\documentclass[twocolumn,prl,aps,amsfonts]{revtex4}
\usepackage{graphicx}
\usepackage{dcolumn}
\usepackage{bm}
\def\vec#1{\mbox{\boldmath $#1$}}
\begin{document}
\preprint{arXiv:cond-mat/0108192}
\title{Collisional Granular Flow as a Micropolar Fluid}
\author{Namiko Mitarai$^1$, Hisao Hayakawa$^2$, and Hiizu Nakanishi$^1$}
\affiliation{$^1$Department of Physics, Kyushu University 33, 
Fukuoka, 812-8581, Japan\\
$^2$Graduate School of 
Human and Environmental Studies, Kyoto University,
Sakyo-ku, Kyoto 606-8501, Japan}
\date{\today}
\begin{abstract}
We show that a micropolar fluid model 
successfully describes collisional granular flows on a slope.
A micropolar fluid is the fluid with internal structures
in which coupling between the spin of each particle
and the macroscopic velocity field is taken into account.
It is a hydrodynamical framework suitable for
granular systems which consists of particles with
macroscopic size.
We demonstrate that the model equations
can quantitatively reproduce the velocity and the angular velocity profiles 
obtained from the numerical simulation of the collisional granular flow
on a slope
using a simple estimate for the parameters in the theory.
\end{abstract}
\pacs{45.70.Mg, 47.50.+d, 47.85.-g}
\maketitle
In spite of the long history of research
on the granular flows,
the theoretical framework for their rheology has not yet been established.
One factor that makes
analytical treatment difficult
is that there is not a great
separation of the length scales;
the size of each particle is often
comparable with the scale of the macroscopic collective motion.
Therefore, there are many situations in which 
simple hydrodynamic approaches
cannot be used to characterize granular flows
\cite{K99}.
Even when we consider the rapid 
granular flows \cite{rapid}, in which
the density is low enough that kinetic theory 
seems to be valid, the coupling
between the rotation of each particle
and macroscopic velocity field may not be
negligible. Thus, the behavior of the flow
can deviate from that of a Newtonian fluid.

The micropolar fluid model is a continuum model
to describe a fluid 
that consists of particles with spinning motion \cite{L99}.
The model equations include an asymmetric stress tensor and
a couple stress tensor.
Therefore, the model can be a suitable framework to 
describe granular flows.

Although some research on the application of the micropolar fluid model 
to granular flows has been
carried out \cite{mic},
most work has considered dense granular flows, in which 
the constitutive equations adopted for the stress and the couple stress 
tensors were very complicated. 
Hence it was difficult to interpret the results physically.

In this paper, we apply a micropolar fluid model
to a collisional granular flow.
We adopt a set of constitutive equations that are a
simple and natural extension of those for
a Newtonian fluid.
Note that Newtonian fluid models of rapid granular flows
\cite{rapid}
have been well established.
We calculate the velocity and angular velocity profiles
for uniform, steady flow on a slope 
and demonstrate that the micropolar fluid model reproduces
the results of numerical simulations.

It is easy to derive the equations
for the number density $n$, the velocity $v_i$,
and the microrotation field $\omega_i$
of a system that consists of identical particles 
with mass $m$ and moment of inertia $I$.
From the conservation laws of mass, 
momentum, and angular momentum \cite{L99}, we obtain
\begin{eqnarray}
D_t n&=&-n \partial_k v_k,
\label{eq:ncons}\\
mn D_t v_i
&=&mn f_i+\partial_jS_{ji},
\label{eq:mcons}\\
In D_t
\omega_i
&=&\partial_j C_{ji}
+s^{(a)}_i.
\label{eq:angle}
\end{eqnarray}
The summation convention applies to 
repeated subscripts.
$\partial_i$ represents a partial
derivative with respect to the $i$th coordinate,
$D_t\equiv \partial/\partial t+v_k\partial_k$
is Lagrange's derivative, and
$f_i$ is the body force per unit mass.
Here, $S_{ij}$ and $C_{ij}$ are
the stress tensor and the couple stress tensor
that, respectively, represents the $j$ component of the 
surface {\it force} and the surface {\it torque} 
acting on the plane perpendicular 
to the $i$ axis per unit area, and
$s^{(a)}_i$ is the torque due to the
asymmetric part of the stress tensor defined as
\begin{equation}
s^{(a)}_i=\epsilon_{ijk}S_{jk},
\label{eq:sa}
\end{equation}
where $\epsilon_{ijk}$ is the alternating tensor of Levi-Civita.

For the constitutive equation of the stress tensor $S_{ij}$, 
we adopt \cite{L99}:
\begin{eqnarray}
S_{ij}&=&(-p+\lambda \partial_k v_k)\delta_{ij}
+\mu(\partial_i v_j+\partial_j v_i)\nonumber\\
&&+\mu_r\left[(\partial_iv_j-\partial_j v_i)-2\epsilon_{ijk}\omega_k
\right],\label{eq:sigma}
\end{eqnarray}
with $\delta_{ij}$, Kronecker's delta. 
The symmetric part of $S_{ij}$ in Eq. (\ref{eq:sigma})
is the same as the stress tensor of the Newtonian fluid
with pressure $p$,
shear viscosity $\mu$, and bulk viscosity $\lambda$.
The third term on the right hand of 
Eq. (\ref{eq:sigma}) represents the asymmetric part
of the stress tensor due to
the difference between the rotation of the mean velocity field and
particles' own spin. This gives
$\vec{s}^{(a)}=2\mu_r[\nabla\times\vec{v}-2\vec{\omega}]$,
using Eq. (\ref{eq:sa}).
The microrotation viscosity $\mu_r$ 
represents the coupling between the velocity and the microrotation field.
For the couple stress tensor $C_{ij}$,
we use the theorem that isotropic second order tensors
can be expanded as the trace,
the symmetric part, and the asymmetric part 
of rate of strain tensor for microrotation \cite{L99}:
\begin{eqnarray}
C_{ij}&=&c_0\partial_k\omega_k\delta_{ij}
+\frac{\mu_B+\mu_A}{2}(\partial_i\omega_j+\partial_j\omega_i)\nonumber\\
&&+\frac{\mu_B-\mu_A}{2}(\partial_i\omega_j-\partial_j\omega_i),
\label{eq:couple}
\end{eqnarray}
where the coefficients of angular viscosity are $c_0$, $\mu_A$, and
$\mu_B$.
It should be noted that the dimensions of these coefficients are
different from the viscosities in the stress tensor by
length-squared because of the difference of the dimensions
between $S_{ij}$ and $C_{ij}$ and between 
$v_i$ and $\omega_i$.

It is debatable whether such a straightforward extension
of Newtonian constitution
relations can be applied to granular flow, because 
the hydrostatic term
in a granular material
should have a different form.
For a collisional flow, however,
we expect that such an effect is not important.
Therefore, we concentrate on a collisional granular flow.

The coefficients of viscosity that appear in Eqs. 
(\ref{eq:sigma}) and (\ref{eq:couple}) have been
derived based on the kinetic theory of three-dimensional
spheres with rough surfaces \cite{vis1,vis2}.
Here, for later convenience, and also to make the physical
meaning of the model clear, we briefly summarize the 
rough estimate of the coefficients of viscosity
to the lowest order
for two-dimensional disks using 
elementary kinetic theory.
Let us consider a two-dimensional fluid
that consists of identical disks with 
diameter $\sigma$ 
and that is flowing uniformly in the $x$ direction, namely
$n=n(y)$,
${\vec v}=(u(y), 0, 0)$, and
${\vec \omega}=(0, 0, \omega(y))$.
Then we have
\begin{equation}
S_{yx}=\mu u'(y)+\mu_r[u'(y)+2\omega(y)],
\label{eq:sigmayx}
\end{equation}
and 
\begin{equation}
C_{yz}=\mu_B\omega'(y),
\end{equation}
where the prime denotes a derivative with respect to its argument.
Here, $S_{yx}$ ($C_{yz}$) is the $x$ ($z$) component of the force 
(torque) per unit area acting 
on the plane perpendicular to the $y$ axis.

The coefficient $\mu$ in Eq. (\ref{eq:sigmayx}) 
corresponds to the kinetic viscosity in
dilute gas, which
we can find an estimate of kinetic theory in textbooks 
on statistical physics, e.g. Ref. \cite{Reif}.
It is given by
\begin{equation}
\mu\sim n\bar v m l\sim \frac{1}{\sigma}\sqrt{Tm},
\end{equation}
where $l$ is the mean free path
and $\bar v$ is the mean square displacement
of the velocity in the $y$ direction.
Here, $T$ is the granular temperature
and the relation $l\sim 1/(n\sigma)$ 
in two-dimensional system 
is used.

The coefficient $\mu_B$ is estimated by 
a similar argument to that of $\mu$.
Because $C_{yz}$ represents the net angular momentum transfer
per unit time per unit length
due to the microrotation
by particles crossing the plane $y=\mbox{const}$,
we can use the argument for $\mu$
by replacing $u(y)$ and $m$ by $\omega(y)$ and $I$, respectively.
Then $\mu_B$ is estimated as
\begin{equation}
\mu_B\sim n\bar v I l\sim \sigma\sqrt{Tm},
\end{equation}
with $I=m\sigma^2/8$ for a two-dimensional disk.
As we have mentioned, the dimensions of $\mu$ and $\mu_B$
are different.

For an estimation of $\mu_r$,
which gives coupling between particles' own spin 
and velocity field,
we consider the collision of two disks 
near the plane $y=\mbox{const}$ \cite{comment1}.
If the surface of the disk has some roughness,
the momentum tangent to the 
relative position of the colliding particles
at the time of contact
is transferred from one particle to another.
It is plausible to assume 
that the tangential momentum transferred in one collision
is proportional to $m$ times $\Delta u$,
the relative tangential 
velocity of each particle at the contact point. 
In order to simplify the estimation,
let us consider the situation with 
a uniform velocity field, namely
$u'(y)=0$ \cite{comment2}.
Then $\Delta u$ is given by
$\Delta u \sim 2(\sigma/2)\omega(y)$.
Because the frequency of collision per unit time
per unit length near the plane $y=\mbox{const}$ with the width $\sigma$
is proportional to $n^2\bar v \sigma ^2$ in two-dimensions,
the momentum transfer
across the plane by collisions is estimated as
\begin{equation}
\Delta M\sim n^2\bar v \sigma^3m[2\omega(y)].
\label{eq:delM}
\end{equation}
Comparing Eq. (\ref{eq:delM}) and the second term of
Eq. (\ref{eq:sigmayx}) with $u'(y)=0$,
we obtain
\begin{equation}
\mu_r\sim n^2 \bar v \sigma^3 m
\sim n^2\sigma^3\sqrt{Tm}.
\label{eq:mur}
\end{equation}

Summarizing the results above, which are
consistent with Ref. \cite{vis1}, we get following expressions
for the coefficients of viscosity;
\begin{equation}
\frac{\mu_B}{\mu_r}\sim\frac{1}{n^2\sigma^2}\sim l^2,
\quad 
\frac{\mu}{\mu_r}\sim\frac{1}{(n\sigma^2)^2}
\sim \left(\frac{l}{\sigma}\right)^2.
\label{mumu}
\end{equation}
It should be noted that, because the dimension of $\mu_B$
is different from that of $\mu$ and $\mu_r$,
we need to introduce another length scale to
characterize the macroscopic flow behavior in order to compare them.
On the other hand, we can see that, 
when the number density is high enough,
$\mu_r$ becomes comparable to $\mu$, then
the coupling between the angular momentum
and the linear momentum should play an important role.

Now we present the 
uniform, steady solution of the micropolar fluid equations
(\ref{eq:ncons}), (\ref{eq:mcons}), and (\ref{eq:angle}) on a slope,
and compare the obtained profiles with the result of 
numerical simulation.
\begin{figure}
\begin{center}
\includegraphics[angle=-90,width=4.cm]{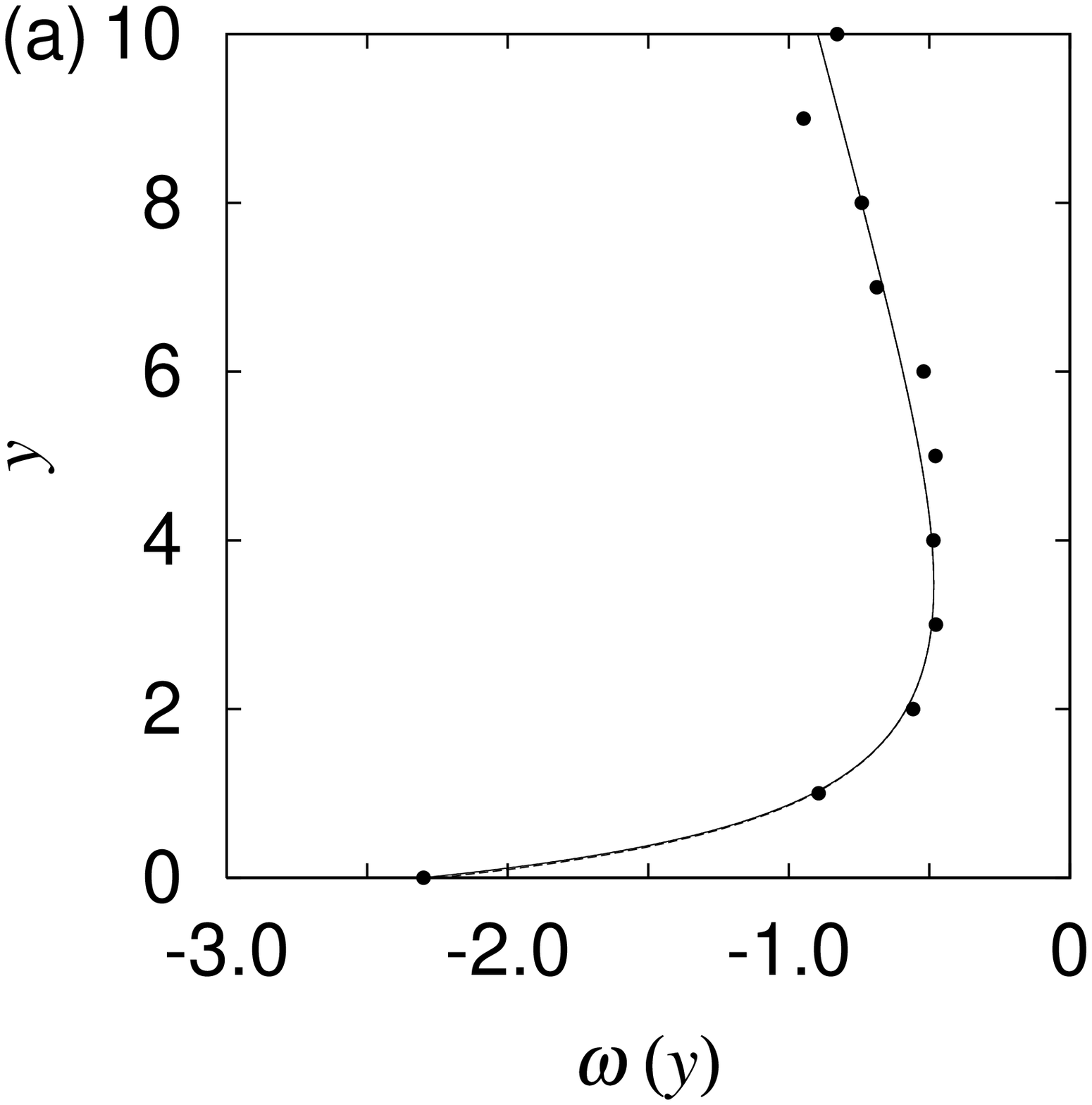}
\includegraphics[angle=-90,width=4.cm]{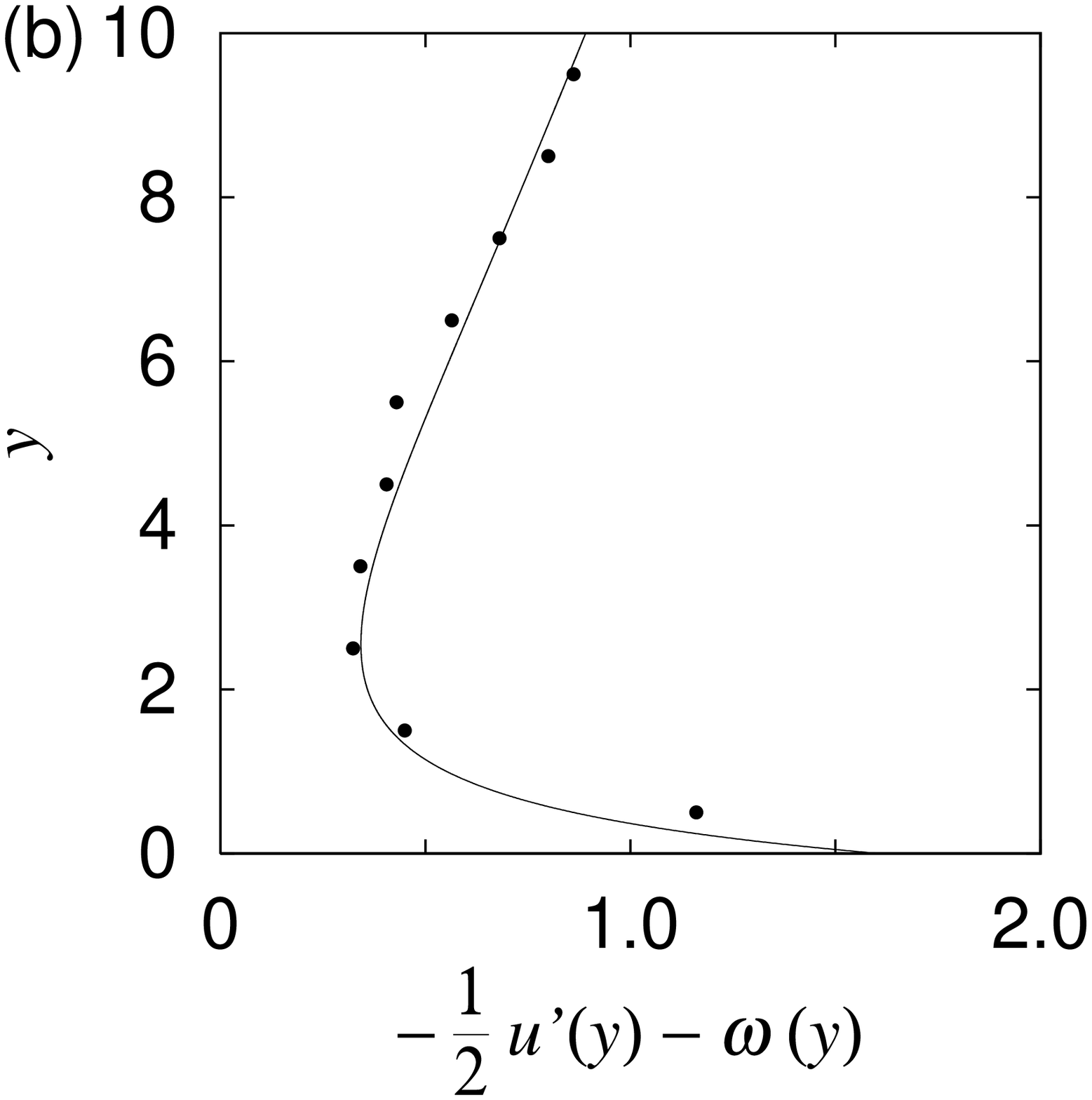}
\end{center}
\caption{The properties of the uniform flow. Filled circles show
the data from the simulation and the solid line shows
the uniform, steady solution of the micropolar fluid equations.
(a) The angular velocity profile $\omega(y)$.
The dashed line is the 
finite series approximation of the solution Eq. (\ref{eq:omsol})
with $k=1$ to $5$; the difference from the solid line is 
hardly distinguished by the eye.
(b) The profile of the deviation of 
the microrotation from
the rotation of velocity field,
$\frac{1}{2}\nabla \times {\vec v}-{\vec \omega}$.
}
\label{fig}
\end{figure}

Let us consider the two-dimensional steady flow on 
a slope under the gravity.
We take the $x$ axis in the direction down the slope,
and the $y$ axis in the direction perpendicular to the slope.
The inclination angle of the slope is $\theta$,
and the acceleration of gravity ${\vec g} =(g\sin\theta, -g\cos\theta, 0)$.
In the uniform, steady flow, $mn=\rho=\rho(y)$,
${\vec v}=(u(y), 0, 0)$, and ${\vec \omega}=(0, 0, \omega(y))$,
therefore the equation of continuity (\ref{eq:ncons})
is automatically satisfied.
Eqs. (\ref{eq:mcons}) and (\ref{eq:angle}) reduce to following 
differential equations:
\begin{eqnarray}
\rho g\sin\theta+
\frac{\mbox{d}}{\mbox{d}y}
\left[\mu
\frac{\mbox{d}u}{\mbox{d}y}+
\mu_r\left(
\frac{\mbox{d}u}{\mbox{d}y} +2\omega\right)
\right]&=&0,
\label{eq:mx}\\
-\rho g\cos\theta-
\frac{\mbox{d}p}{\mbox{d}y} &=&0,
\label{eq:my}
\end{eqnarray}
and
\begin{equation}
-2\mu_r\left(
\frac{\mbox{d}u}{\mbox{d}y} 
+2\omega\right)
+\frac{\mbox{d}}{\mbox{d}y}
\left[\mu_B\frac{\mbox{d}\omega}{\mbox{d}y}
\right]=0.
\label{eq:ang}
\end{equation}
The equation of state for a two-dimensional granular gas 
has been derived in Ref. \cite{JR85},
but here we adopt the lowest order estimate, 
namely $p=\rho T/m$.
With the aid of the assumption of constant temperature
$T=\bar T$,
we obtain the density profile from Eq. (\ref{eq:my});
\begin{equation}
\rho=\rho_0\exp\left(-\frac{y}{h}\right),\
h=\frac{\bar T}{mg\cos\theta}.
\label{eq:denspro}
\end{equation}
Then, from the estimate of the coefficients Eq. (\ref{mumu}),
we introduce the non-dimensional constants
$\alpha$ and $\beta$ as
\begin{equation}
\frac{\mu_r}{\mu}=\alpha\exp(-2y/h),
\quad
\frac{\mu_B}{\mu}=\beta \sigma^2.
\label{eq:param}
\end{equation}
After integrating Eqs. (\ref{eq:mx}) and (\ref{eq:ang})
with Eqs. (\ref{eq:denspro}) and (\ref{eq:param}),
we obtain the following relation between the velocity field 
$u(y)$ and the microrotation field $\omega(y)$;
\begin{equation}
\tilde u(Y)=-\exp(-Y)-\frac{\beta\epsilon^2}{2}
\frac{\mbox{d}\tilde\omega(Y)}{\mbox{d}Y}+A_0,
\label{eq:utilde}
\end{equation}
where
\begin{equation}
\tilde u
=\frac{u}{\rho_0 g h^2\sin\theta/\mu},\quad
\tilde \omega=\frac{\omega}{\rho_0 g h\sin\theta/\mu},
\end{equation}
with $Y=y/h$ and $\epsilon=\sigma/h$.
Here, $A_0$ is an integration constant.
In the integration of Eq. (\ref{eq:mx}),
we imposed the boundary condition for stress tensor
at the free surface, $\lim_{y\to\infty}S_{yx}\to 0$.
From Eqs. (\ref{eq:ang}) and (\ref{eq:utilde}),
we have
\begin{equation}
\frac{\mbox{d}^2\tilde \omega (Y)}{\mbox{d}Y^2}
-2\frac{\alpha \exp(-2Y)}{\beta\epsilon^2}
\left[\frac{\exp(-Y)+2\tilde \omega(Y)}
{1+\alpha\exp(-2Y)}\right]=0.\label{eq:omdifY}
\end{equation}
Its general solution is given as
the sum of a particular solution $\tilde \omega_p$
and two homogeneous solutions $\tilde \omega_1$ and $\tilde \omega_2$ by
\begin{equation}
\tilde\omega(Y)=
A\tilde \omega_1(Y)+B\tilde \omega_2(Y)+
\tilde \omega_p(Y),
\label{eq:omsol}
\end{equation}
with integration constants $A$ and $B$.
Changing the variable from $Y$ to
$\eta=\exp(-Y)$, we can obtain the expressions
\begin{eqnarray}
\tilde\omega_p&=&\sum_{k=1}^{\infty} a_{2k+1}\eta^{2k+1},
\quad \mbox{with} \quad 
a_3=\frac{2\alpha}{9\beta\epsilon^2},  \nonumber\\
a_{2k+1}&=&f_{2k+1}a_{2k-1},\\
\tilde\omega_1&=&\sum_{k=0}^{\infty} b_{2k}\eta^{2k}, \quad 
\mbox{with} \quad b_0=1, \nonumber\\
b_{2k}&=&f_{2k}b_{2k-2},
\end{eqnarray}
and 
\begin{eqnarray}
\tilde\omega_2&=&\tilde\omega_1\log\eta+\sum_{k=1}^{\infty} c_{2k}\eta^{2k}, 
\quad \mbox{with} \quad c_0=0,\nonumber\\
c_{2k}&=&f_{2k}c_{2k-2}
-\frac{\alpha}{k^3}
\left[\frac{1}{\beta\epsilon^2}+k-1\right]b_{2k-2},
\end{eqnarray}
where
\begin{equation}
f_m\equiv\frac{\alpha}{m^2}\left[
\frac{4}{\beta\epsilon^2}-(m-2)^2\right].
\end{equation}
Note that $\tilde \omega_p$ and $\tilde \omega_1$ are regular
but $\tilde \omega_2$ diverges at $\eta=0$ ($Y\to \infty$).

Now we show that the solution
obtained from the micropolar fluid model
can quantitatively reproduce the results of numerical simulations
of collisional granular flow on a bumpy slope \cite{MN01}.
In the simulation, the discrete element method is employed
with normal and tangential elastic force and dissipation.
The particles are modeled by disks with diameter $\sigma$,
and the parameters are chosen so that the normal restitution
$e_n$ becomes $0.7$.
The surface of the slope is made rough by attaching
identical particles to it.
In the following, all quantities are given in dimensionless form 
with mass unit $m$, length unit $\sigma$, and time unit $\tau=\sqrt{\sigma/g}$.

We used the data of the simulation with inclination
$\sin\theta=0.45$, system size $L=1002$
and number of flowing particle $N=1000$.
In the simulations, uniform flow is realized during
$500 \lesssim t\lesssim 2000$ where $t$ is the time \cite{MN01}.
In order to determine the profiles of the mean quantities 
describing the flow, 
we divide the space into layers which are one particle diameter wide 
parallel to the slope, calculate the averages inside layers,
and then average the data over the time 
within the uniform flow, $1000\le t < 1500$.
The origin $y=0$ is 
taken one diameter above from the top of the disks attached to the slope.

Now let us compare our analytical result from the micropolar fluid model
with the simulation data.
First, we confirmed that the number density profile
can be well-fitted by Eq. (\ref{eq:denspro}) with $h=2.24$, namely
$\bar T\sim 2.0$, which has been checked to be
close to the averaged value of $T$. 
In order to fit the solutions (\ref{eq:utilde})
and (\ref{eq:omsol}) to the simulation data,
we determine the boundary values $u(0)$ and $\omega(0)$
from the data
and treat $\alpha$, $\beta$, $\mu/\rho_0$ 
and $\omega'(0)$ as fitting
parameters.
From Fig. \ref{fig} (a),
we can see that the micropolar fluid equations
reproduce the angular velocity quantitatively.
Here, the value of parameters are
$\alpha\sim 0.10$, $\beta\sim 0.12$, 
$\mu/\rho_0\sim 0.95$, and $\omega'(0)\sim 2.9$.
The velocity profile $u(y)$ can also be well-reproduced
by the solution (\ref{eq:utilde});
however, because the density is low,
the deviation of the velocity profile from the 
solution of the Navier-Stokes equation, i.e. 
$\tilde u(Y)=-\exp(-Y)+A_0$, is small.
Actually, we have checked that the data can be reasonably fit
with any small value of $\alpha$, as long as
$\alpha/\beta$ is chosen appropriately.
However, it is significant that the solution (\ref{eq:omsol})
can reproduce the sharp variation of $\omega(y)$ 
over only a few diameters near the base.

The more important thing is that 
the mean spin $\vec{\omega}$ deviates systematically
from the rate of bulk rotation
$\frac{1}{2}\nabla \times \vec{v}$, and that
the micropolar fluid model can reproduce this deviation.
In Fig. \ref{fig} (b), we see that the deviation is
large near the base,
because each particle is forced to rotate by the collision with the slope. 
This result indicates the importance of the 
couple stress near the boundary, 
as pointed out in Ref. \cite{cam}.
This deviation may produce 
the velocity profile different from the 
Newtonian fluids described by the Navier-Stokes equation
due to the coupling between the spin of each particle
and the linear velocity field \cite{J92}.
On the other hand, it seems the deviation
also becomes large
in the region far from the boundary.
The reason is that 
the microrotation field in this region 
is dominated by a small number of particles spins
that are generated  by collisions
with the boundary.

In summary, 
the micropolar fluid model has been applied
to a collisional granular flow with relatively low density.
It has been demonstrated that 
the solution for uniform, steady flow on an inclined surface
reproduces the results of numerical simulation.

Because the density is low,
the ideal gas assumption for the equation of state and the estimation of the
viscosity from the elementary kinetic theory work well.
In order to apply the model to denser collisional flows,
we need a systematic extension of the theory using, 
for example, Enskog theory.
This has already been done in the context of
polyatomic fluids
for completely rough spheres without energy dissipation
\cite{vis1} and it should be possible to extend such results to the
dissipative case \cite{vis2}.
The equation of state for dense granular gas 
has also been discussed recently \cite{state}.
On the other hand, it is known that
the effect of the particle spin becomes more
important in dense frictional flows \cite{dense}.
Much research has been done to construct 
the mechanics of granular media which is valid 
not only for collisional flow but also for denser situations 
based on the micropolar or Cosserat theory \cite{densemicro}.
The concepts of micropolar mechanics may
help to link the understandings of the collisional 
and the frictional flows of granular materials. 

The authors are grateful to J. T. Jenkins 
for a careful reading of the manuscript.
N. M. and H. H. also thank him and A. Shimosaka 
for fruitful discussions.
N. M. and H. N. are grateful for the warm hospitality during their stay at
YITP of Kyoto University where part of this work
has been carried out.


\begin{thebibliography}{99}
\bibitem{comment1}
It should be noted that, 
such a collision process produces the (angular) momentum transfer 
related to not only $\mu_r$ but also $\mu$ ($\mu_B$).
In this paper, only the lowest order in {\it each}
viscosity is considered, therefore such 
corrections to $\mu$ and $\mu_B$ are neglected.
\bibitem{comment2}
If we consider the situation with $u'(y)\ne 0$, 
we get $\Delta u=\sigma[u'(y)\cos^2\theta+\omega(y)]$,
where $\theta$ is the angle between the $y$ axis and
the vector along the centerline from 
one particle to another.
We need careful calculation in the average process of $\theta$
which is beyond elementary kinetic theory.
We confirmed that a more systematic derivation of $\mu_r$ 
similar to the method in Ref. \cite{vis2}
gives the asymmetric part of the stress tensor 
proportional to $u'(y)+2\omega$.
\end{thebibliography}
\end{document}